\documentclass[12pt,a4paper]{article}

\usepackage[cp1251]{inputenc} 
\usepackage[T1]{fontenc} 
\usepackage{microtype} 
\usepackage{mathrsfs}
\usepackage[english]{babel} 
\usepackage{bbm}
\usepackage{amsmath,amsfonts,amscd,amssymb,latexsym}
\usepackage{dcolumn}

\usepackage[unicode, pdftex]{hyperref}
\usepackage[usenames]{color}

\usepackage{graphics}
\usepackage{titling}
\usepackage{ textcomp }
\graphicspath{{/}}
\DeclareGraphicsExtensions{.eps,.pdf,.png,.jpg}
\usepackage{wrapfig}
\usepackage{amssymb}
\usepackage{amsmath}
\usepackage[dvips]{graphicx}
\usepackage{rotating}
\usepackage{longtable}
\usepackage{multirow}
\usepackage{geometry}
\usepackage{fancyhdr}
\begin{document}


\newenvironment{sciabstract}{
\begin{quote} \scriptsize}
{\end{quote}}


\title{{\normalsize{\textbf{ADVANCES IN NUCLEAR SCIENCE AND APPLICATIONS}\\
Volume 1, No. 2 (2025), P. 69-79}} \\[2.0cm]
The axion signature of strange quark star in\\ SGR 0501$+$4516}

\author
{C.R. Das\\
\small{The Bogoliubov Laboratory of Theoretical Physics, JINR, Dubna, Russian Federation}\\
\small{e-mail: das@theor.jinr.ru}\\
}

\date{}
\maketitle

\begin{flushleft}
\scriptsize{DOI: \href{https://doi.org/10.63907/ansa.v1i2.35}{10.63907/ansa.v1i2.35}}\\
\scriptsize{Received 10 June 2025}
\end{flushleft}

\begin{abstract}
We study the axion effects on quark matter and quark-matter cores in strange quark magnetars using a three-flavor Nambu-Jona-Lasinio model to represent the charge-parity violating effects through the axion field. Here, axions decay to gamma rays in a very strong magnetic field, which the Fermi Large Area Telescope (Fermi-LAT), Imaging X-ray Polarimetry Explorer (IXPE), and XMM-Newton will be able to detect.


\end{abstract}

\label{Int}
\section{Introduction}

A strange star is a hypothesized compact star that is dense enough to literally break down ordinary neutrons into their constituent quarks. Furthermore, the up and down quarks are squeezed into an even rarer sort of quark known as a strange quark, which explains the name strange star. Technically, up, down, and strange quarks make up the ``strange'' matter of a strange star. This mixture of sub-hadronic particles may be even more stable than a typical neutron star. Strange quark matter could represent the real ground state of dense matter, a mixture of almost equal amounts of deconfined up, down, and strange quarks \cite{int1,int2}. It is possible that the detected pulsars are strange quark stars, also known as strange stars, if the strange quark matter hypothesis is accurate. Strange quark stars (SQS) offer perfect natural laboratories for studying physics under harsh astrophysical settings because of their extremely high densities, powerful electromagnetic fields, and strong gravitational fields \cite{int3,int4,int4a}.

Large ambiguities in the internal structure of strange stars result from the lack of clarity surrounding the nature of the strongly interacting materials at such extreme densities. Theoretical and observational features of strange stars have been the subject of much study, yet there are still many unanswered questions. Strong quark-quark interactions inevitably self-bind strange quark stuff. Therefore, compact stars made of strange quark matter may be bare strange stars with a surface density that suddenly drops to zero. But regular nuclear materials can likewise make up a thin crust of a peculiar star \cite{int5}. The interior structure of strange stars is hardly affected by the light crust since the maximum density of the crust is five times lower than the neutron drip density \cite{int6,int7}. On the other hand, it might drastically alter the properties of electromagnetic emissions from these tiny stars. Additionally, certain types of electromagnetic explosions or even the release of gravitational waves could result from the crust collapsing.
 
Magnetars are a type of rare neutron star with a magnetic field that is the most powerful in the universe, approximately a thousand times stronger than that of a typical neutron star and a quadrillion times stronger than that of Earth. As a spinning magnetar can progressively collapse into an even more dense form through glitches of inside vortices, which would be something akin to a strange star with the requisite mix of quarks. It would undoubtedly cause gamma-ray and X-ray outbursts in near-infrared (NIR) imaging of soft gamma repeater (SRG) highly magnetized neutron stars \cite{int8,int9,int10}. We found SGR 0501$+$4516 is a magnetar renowned for its gamma-ray and X-ray bursts and is a candidate for a strange quark star.

We use a three-flavor Nambu-Jona-Lasinio (NJL) model \cite{int10a} and similar ideas \cite{int11} to describe the charge-parity violating effects through the axion field in order to examine the axion impacts on quark matter and quark-matter cores in strange quark magnetars. In this scenario, a photon vertex is where axions decay to gamma rays. We can use the theoretical emissivity to calculate the predicted flows. These emissions can be detected with the Fermi Large Area Telescope (Fermi-LAT), Imaging X-ray Polarimetry Explorer (IXPE), and XMM-Newton.

\label{NJLModel}
\section{NJL Model}

Nambu and Jona-Lasinio \cite{1,2} first presented the NJL model to explain the interaction of nucleons. Eguchi and Sugawara \cite{3} expanded it to encompass up and down quarks. The idea was expanded by Kikkawa \cite{4} to include up, down, and strange quarks, among other quark flavors. In this review, we characterize quark matter inside strange stars using the three-flavor NJL model. In order to comprehend the enormous nucleon mass and the dynamic creation of fermion masses, the model demonstrates spontaneous violation of chiral symmetry. The model is also notable for being solvable, which in some limiting instances enables the acquisition of straightforward analytical results.

The three-flavor NJL model with axion fields' Lagrangian density is  \cite{12}:
\begin{eqnarray}\label{eq1}
\mathcal{L}&=& \bar{\psi}\left(i\rlap{\slash}\partial-\hat{m}\right)\psi\notag\\
&&+\frac{G_{S}}{2}\displaystyle\sum_{a=0}^{8}\left[\left(\bar{\psi}\lambda_{a}\psi\right)^{2}+\left(\bar{\psi}i\gamma_{5}\lambda_{a}\psi\right)^{2}\right] 
\notag\\
&&-\frac{G_{V}}{2}\displaystyle\sum_{a=0}^{8}\left[\left(\bar{\psi}\gamma_{\mu}\lambda_{a}\psi\right)^{2}+\left(\bar{\psi}\gamma_{5}\gamma_{\mu}\lambda_{a}\psi\right)^{2}\right] \notag\\
&&- K\left\{e^{i\frac{a}{f_a}}\det\left[\bar{\psi}\left(1+\gamma_{5}\right)\psi\right]+e^{-i\frac{a}{f_a}}\det\left[\bar{\psi}\left(1-\gamma_{5}\right)\psi\right]\right\}.
\end{eqnarray}

The quark field with three flavors is represented by $\psi = (u, d, s)^T$, the current quark mass matrix is $\hat{m} = diag(m_u,m_d,m_s)$, and the flavor SU(3) Gell-Mann matrices are represented by $\lambda_a$, where $\lambda_0 = \sqrt{2/3}I$. The scalar and vector coupling constants are $G_S$ and $G_V$, respectively. The six-point contact that disrupts the axial symmetry $U(1)_A$ \cite{5} is represented by the $K$ term in Eq.~\ref{eq1}. $a(x)=\theta(x) f_a$ is the normalized axion field in the term $K$, where $f_a$ is the axion's decay constant, which also serves as a representation of the Peccei–Quinn (PQ) symmetry \cite{5a,5b,5c,5d,5e} breaking scale. Astrophysical evidence, such as the black hole superradiance and the cooling rate of the SN1987A supernova, heavily constrains the PQ symmetry-breaking scale with $10^8 \leq f_a \leq 10^{17}$ GeV \cite{6,7,8}. Compared to the axion symmetry-breaking energy, we are working with a far smaller energy scale. The interaction between the axion field and the QCD gauge field can therefore be written as $\mathcal{L}_\theta \sim (a/f_a)\text{Tr}\, G_{\mu\nu} \widetilde{G}^{\mu\nu}$. This is because we can assume that the axion field $a$ is in its vacuum expectation value, and for $a/f_a$, it can be within the range of 0 to $\pi$. Here, we use the cutoff value in the momentum integral $\Lambda = 750$ MeV provided in Refs.~\cite{8a,8b} , together with the constants $m_u = m_d = 3.6$ MeV, $m_s =$ 87 MeV, $G_S\Lambda^2 =$ 3.6, and $K\Lambda^5 =$ 8.9 with $G_S=2G_V$.

Quark confinement is frequently utilized in conjunction with the bag constant (B) to derive the EOS because it is not immediately reflected in the NJL model. In contrast to the MIT bag model, the NJL model's confinement is a dynamical result rather than a manual imposition. But just like the MIT bag model, the NJL model is unable to account for the properties of asymptotic freedom. Furthermore, this model does not include gluons and instead views quark-quark interaction as point-to-point interaction. It is not a renormalizable field theory as a result, and in order to deal with the improper integrals that emerge, a regularization method needs to be given \cite{9}. Readers can consult the review articles by Buballa \cite{10} and Klevansky \cite{9} for further information and also \cite{11,12}.

\label{SGR0501+4516}
\section{SGR 0501$+$4516 the Strange Quark Magnetar}

SGR 0501$+$4516 is a magnetar, a rare kind of neutron star that has a magnetic field that is enormous ($2\times 10^{15}$ Gauss). These are its salient characteristics: A soft gamma repeater (SGR) repeatedly emits gamma or X-ray bursts. During a significant eruption in August 2008, NASA's Swift satellite picked it up. It is 16,000 light-years away from Earth and is part of the constellation of Camelopardalis (Giraffe). Its spin-down rate is a rapid slowing brought on by magnetic braking, and its rotation period is 5.76 seconds \cite{13,14}. Detected for the first time with strong gamma-ray bursts in August 2008 and released fresh X-ray blasts in 2016 following an 8-year hibernation \cite{15}. We assumed it is a strange quark magnetar.

A definitive unification of SGRs, anomalous X-ray pulsars (AXPs), and transient AXPs (TAXPs) into a single class of ``magnetar candidates'' has been proposed, citing SGR 0501$+$4516 and 1E 1547.0-5408 as potential tools \cite{15x}. No supernova remnants were discovered in the region where SGR 0501$+$4516 had passed through throughout its anticipated lifetime, according to a 2025 analysis. It came to the conclusion that some magnetars may be much older than anticipated, that their progenitors generate low supernova ejecta masses, or that they may develop through low-mass neutron star mergers or accretion-induced collapse \cite{15y}.

\label{AEQS}
\section{Axion Emissivity of the Strange Star}

The QCD axion \cite{15z} has been studied since its emergence in nucleon-nucleon bremsstrahlung in supernovae or neutron stars. Many neutron star models incorporate QCD phases \cite{15a,15b,15c}, which include both hadronic and quark matter. The neutron star may be in a mixed phase of hadronic and quark matter or a QCD phase of quark matter in the scenario we consider \cite{16}. This quark matter phase is believed to be color superconducting. Furthermore, the color-flavor locking (CFL) phase, in which strange quarks participate in the Cooper pairing mechanism with up and down quarks, can be taken into consideration \cite{17}. This phase has broken chiral symmetry and is superfluid. Stellar matter candidates include the gapless two-flavor color superconducting (g2SC) phases, the gapless CFL, and the neutral quark (NQ) \cite{15a}.

In addition to serving as a valuable tool for quantifying the energy released by strange quark star axions, the emissivity can be used as a bridge to compute the flux, gamma-ray spectral energy distribution (SED), and gamma-ray energy flow. The axion-quark coupling, which can be expressed using the form factors and the Goldberger-Treiman relation, allows us to first determine the axion emissivity \cite{19}. It could be more appropriate to examine the axion coupling to quarks via the triangle diagram and the chiral anomaly mediated by gluons in the model of QCD phase of matter for strange quark stars. One way to describe the axion-to-gluon connection is as follows:

\begin{equation}
{\cal L}\subset\frac{a}{f_a} G_{\mu\nu}\widetilde{G}^{\mu\nu}.
\end{equation} 

We derive the quark couplings to axions $g_{au\bar u}$, $g_{ad\bar d}$, and $g_{as\bar s}$ using the axial vector current \cite{19a}. It will be demonstrated that the axion coupling to quarks can meet the requirements of ultralight axions (ULA) and can be more sensitive to lower masses than the nucleon coupling to axions. It is hypothesized that ULA is a significant type of dark matter \cite{19b}.

The Goldberger-Treiman relation can be used to express the axion-quark-antiquark coupling:
\begin{equation}
g_{aq\bar{q}}\left(q^2_s\right)=\frac{q^2}{2f_a}\left|F_3^5\left(q^2\right)\right|.
\end{equation} 
Where the tree-level coupling's form factor can be expressed as:
\begin{equation}
\left|F_3^5\right|=\left(\frac{1}{q^4}f_a^2\alpha_s^4\right)^{-1}.
\end{equation}
For tree-level coupling, we use $q_s=87$ MeV, $f_a\simeq 10^{15}$ GeV, $\alpha_s\simeq 10^{-2}$, and $F^5_3\simeq 10^{-26}$. When thinking about the coupling of quarks to axions, it is reasonable to ignore the tree-level s-channel (by gluon exchange) coupling. In this scenario we do not include one-loop coupling because it affects the axial vector current, and furthermore, $F^5_3|_{\rm tree}\ll F^5_3|_{\rm two-loop}$.

The amplitude of the cross-section can be expressed as follows:
\begin{equation}
 \sigma\left(q\bar{q}\to a\right)=\frac{12\pi^2}{M_q}\Gamma\left(a\to q\bar{q}\right)\delta\left(M_q^2-M_s\right),
\end{equation}
where $q\bar{q}$, $q^2$, $f_a$, $F_3^5$, and $M_s$ are quark/anti-quark state, momentum transferred squared, axion decay constant, form factor, and center-of-mass (CM) energy, respectively \cite{26}.

\label{ipe}
\section{Inverse Primakoff Effect}

\begin{figure}[t]
\begin{center}
\includegraphics[width=50mm]{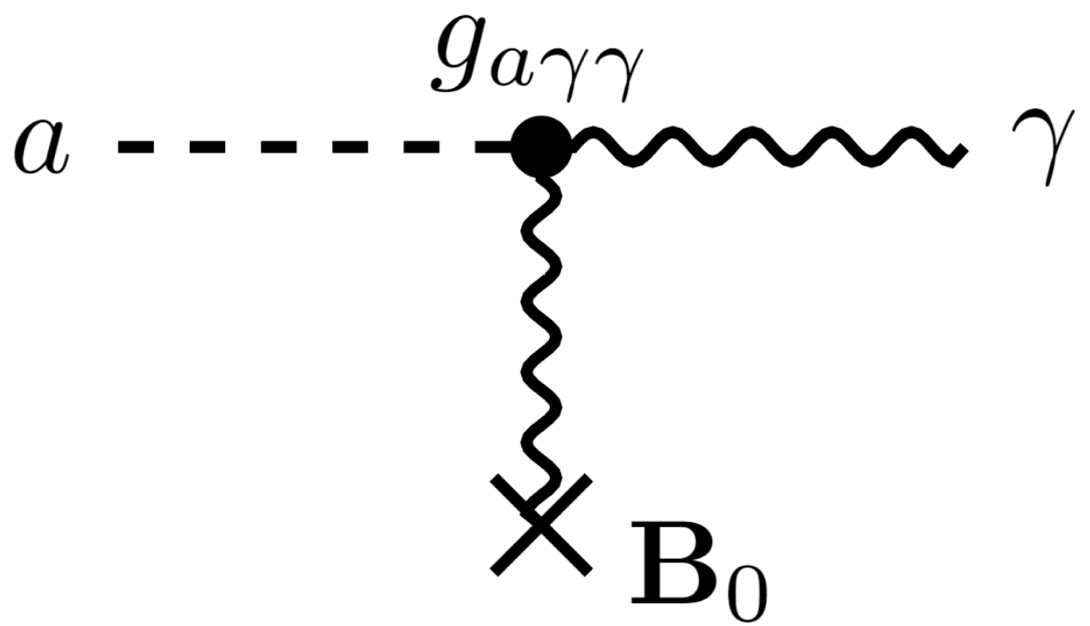}
\includegraphics[width=50mm]{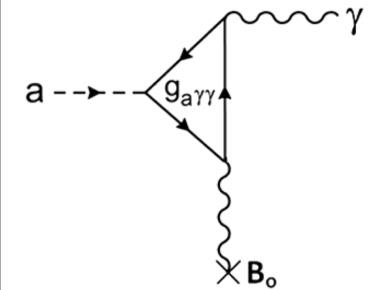}
\vspace{-3mm}
\caption{Tree-level and one-loop Feynman diagrams for an axion transforming into a photon with energy $\omega \simeq 1.5\, m_a \times 10^{13}$ keV in the presence of a static magnetic field $B_0=B_{\rm{SGR\, 0501+4516}}$, where $m_a$ is axion mass. The static magnetic field ($B_{\rm{SGR\, 0501+4516}}$) exhibits the inverse Primakoff effect.}
\label{fig01}
\end{center}
\vspace{-5mm}
\end{figure}

Sikivie presented a detection approach for axion dark matter via the Primakoff effect, wherein a microwave cavity, infused with a high magnetic field, is employed to resonantly amplify the photon production resulting from axion decay \cite{prim1,prim2,prim3}. According to Sikivie's proposal, a meson-like coherent inverse Primakoff process mediates the interaction between an axionic field $(a)$ and a photon through the axion's tree level and two-photon vertex (as shown by Feynman diagrams in Figure \ref{fig01}) \cite{yj}.

Axions are electrically neutral, pseudoscalar bosons, and the interaction Lagrangian links an axion to a pair of photons and has a form \cite{tt}:

\begin{equation}
{\cal L}_{a\gamma\gamma} = -\frac{g_{a\gamma\gamma}}{4}F_{\mu\nu}\widetilde{F}^{\mu\nu}a=g_{a\gamma\gamma}\overrightarrow{E}\cdot \overrightarrow{B},  
\end{equation}
where $F_{\mu\nu} \equiv \partial_\mu A_\nu - \partial_\nu A_\mu$ is the electromagnetic field tensor, $\widetilde{F}$ is the Hodge dual of the field strength tensor $F$, $a$ represents the axion field, $g_{a\gamma\gamma}$ is a dimensional axion-two-photon coupling constant, and $\overrightarrow{E}$ and $\overrightarrow{B}$ are the electric and magnetic fields, respectively. The compact version of the axion-two-photon coupling can be expressed as follows:

\begin{equation}
g_{a\gamma\gamma}=\frac{\alpha}{\pi}\frac{g_\gamma}{f_a}.
\end{equation}

The form of $g_\gamma$ is explained by two widely accepted models \cite{marsh,marsh1}: the Dine-Fischler-Srednicki-Zhitnitskii (DFSZ) \cite{5d,5e} or the ``Grand Unified Theory'' model and the Kim-Shifman-Vainshtein-Zakharov (KSVZ) \cite{5b,5c} or ``Hadronic axions'' model. These models assign values of 0 and -0.97 for KSVZ and 8/3 and 0.39 for DFSZ \cite{marsh2,marsh3} to $g_\gamma$, respectively. Similarly, the axion-two-photon coupling constant $g_{a\gamma\gamma}$ has been given an upper limit of $10^{-11}$ GeV$^{-1}$. Here, $g_{a\gamma\gamma}$ is set as $10^{-18}$ GeV$^{-1}$ for $f_a \simeq 10^{15}$ GeV. In Ref.~\cite{marsh4}, the Primakoff process has been thoroughly calculated for one-loop radiative corrections.

\label{keV}
\section{Photon Flux Calculation from the Conversion in the keV-Band}

For axion to photon creation, we employ the same inverse Primakoff effect in the magnetar's neighborhood. The $\rm{\bf E\cdot B_0}$ term in the Primakoff-decay Lagrangian requires the resonant mode's electric field to be parallel to the static magnetic field and in-phase in neutron or quark stars to detect axion decay \cite{marsh5,marsh6}. A high magnetic field permeates the magnetar, which is utilized to resonantly boost the amount of photons produced by axion decay. When an axion-nucleon and axion-quark interaction is present, the hot, dense core of a quark magnetar, which is mostly composed of quarks and nucleons, becomes a substantial generator of axions with energy in the keV band; see Figure \ref{fig01}.

Cooper-pair breaking may become significant for specific temperature ranges, although nucleon bremsstrahlung dominates the axion production mechanism in the SQS core. When axions have masses below $10^{-13}$ eV, they can also be produced by the electromagnetic fields that surround pulsars \cite{20}. High-density nuclear physics and SQS interior conditions introduce significant systematic errors into both procedures. The axion emission from the SQS core is spherically symmetric as a leading approximation. See \cite{21} for additional discussions.

The leading-order solution to the quantum mechanical time-dependent perturbation theory could be used to calculate the axion-photon conversion probability \cite{22}. The conversion probability in the non-relativistic limit is \cite{22a}

\begin{equation}
P_{a\to\gamma}\approx \frac{\pi\, g_{a\gamma\gamma}^2\, B_{\rm{SGR\, 0501+4516}}^2}{3 m_a}\sqrt{\frac{r^3}{2G\, M_{SQS}}},
\end{equation}
where $M_{SQS}$ is the strange quark star mass. The strange star radius, a benchmark conversion probability as a function of radial distance $r$ in the unit of $r_{\rm{SGR\, 0501+4516}}$. The $r_{conv}$ refers to the axion-photon conversion radius. At the scale $r_{conv} \sim 30\times r_{SRG 0501+4516}$, the conversion probability is half of the highest asymptotic value. Our one-loop radiative corrections, together with the original tree-level picture, reduce the Primakoff cross section in the forward or backward scattering limit by approximately 37\% due to the correcting term $(\alpha/2\pi)$ \cite{marsh4}.

\begin{figure}[t]
\begin{center}
\includegraphics[width=135mm]{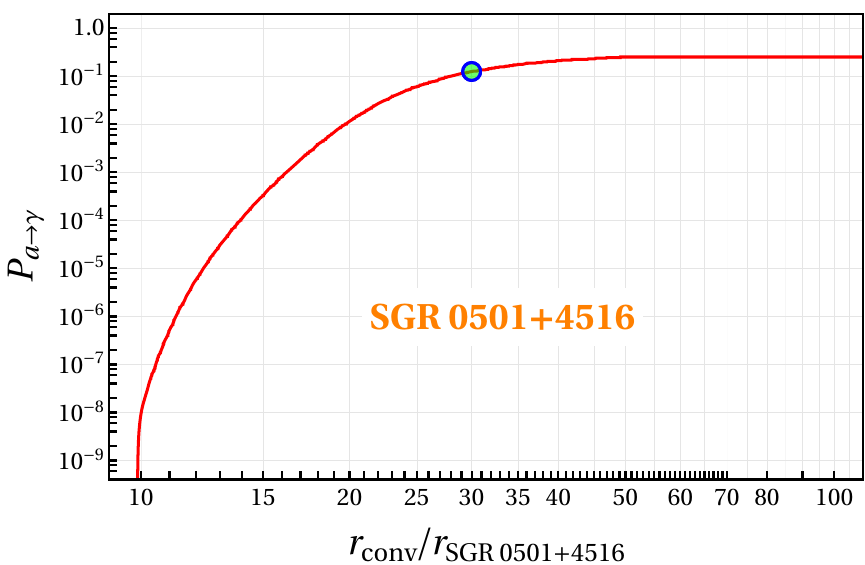}
\vspace{-3mm}
\caption{Axion-to-photon conversion probability $(P_{a\to\gamma})$ as a function of $r_{conv}/r_{\rm{SGR\, 0501+4516}}$ for SGR 0501$+$4516 in Eq.~\ref{eq4}. For the benchmark displayed, it gets close to half of its asymptotic value at $\sim 30\, r_{\rm{SGR\, 0501+4516}}$, which is far from the magnetar surface.}
\label{fig02}
\end{center}
\vspace{-5mm}
\end{figure}

The function $r_{conv}/r_{\rm{SGR\, 0501+4516}}$ for SGR 0501$+$4516 is
\begin{equation}\label{eq4}
\frac{r_{conv}}{r_{\rm{SGR\, 0501+4516}}}=30\times \sqrt[5]{\frac{B(r)^2\, \omega\, r_{\rm{SGR\, 0501+4516}}\, \sin\theta}{\left(2\times 10^{15}\, {\rm Gauss}\right)^2\cdot 5\, {\rm keV}\cdot 10\, {\rm km}}},
\end{equation}
where $\theta$ is the angle between the local magnetic field and the direction of axion (photon) transmission, $\omega$ is the photon energy, and $r_{\rm{SGR\, 0501+4516}} = 10$ km is the usual strange star radius \cite{23}. Only the dipole component of the magnetic field, $B(r) = B_{\rm{SGR\, 0501+4516}}(r_{conv}/r_{\rm{SGR\, 0501+4516}})^{-3}$, is taken into account in the equation above, where $B_{\rm{SGR\, 0501+4516}}$ is the field at the surface. The conversion probability for axions propagating in the strange star magnetosphere is minimal at short radii, as shown in Figure \ref{fig02}, because the mixing is suppressed by the enormous effective photon mass caused by vacuum polarization. The conversion occurs at bigger radii in a dominant manner. As a result, the conversion rate is barely affected by higher-order moments of the magnetic field, which disappear more quickly with radius \cite{24}.

As long as $m_a \lesssim (\omega^2m^2_e/(\alpha r^3_{\rm{SGR\, 0501+4516}}B_{\rm{SGR\, 0501+4516}} \sin \theta))^{1/5}$ \cite{21}, where $m_e$ is the electron mass and $\alpha$ is the fine-structure constant, you will see that $r_{conv}$ is independent of the axion mass $m_a$.

\label{CO}
\section{Conclusions and Outlook\\ as well as Model Predictions}

For all magnetars, this section is dedicated \cite{24a}. There are certain similarities that would be predicted in the energy loss rate, as the calculated emissivity is within several orders of magnitude of the supernova (SN) energy loss rate into neutrinos. Based on this model, we could show a peak in the spectral energy distribution (SED) for $u\bar u$, $d\bar d$, and $s\bar s$ axion coupling for various photon energies extremely close to the Fermi-LAT point source sensitivity of $5\times 10^{-9} {\rm cm}^{-2}\cdot {\rm s}^{-1}$, with a severe cutoff before the Fermi momentum. Although the Fermi-LAT was intended to measure gamma rays with an energy of 30 MeV, this was not feasible in practice. Furthermore, especially for lower temperatures of the strange stars, a multi-messenger analysis involving gamma rays and X-rays would be superior to a Fermi gamma-ray study alone. Since cosmological sub-neV ($\lesssim 10^{-9}$ eV) \cite{marsh,thk} is the lowest mass that is consistent with the physical parameters, we conclude that it is preferred for $m_a$. It would be ideal to detect gamma rays down to 5 MeV if a better gamma-ray telescope were created. Additionally, an X-ray telescope would be complementary to a multi-messenger campaign, monitoring X-ray photons at lower energy \cite{25,26}.

With axions coupling to photons and nucleons and their derived bounds \cite{26a,26b}, the current timing data from XMM-Newton is consistent with the theoretical points selected by the hard X-ray excess, albeit in some disagreement with other astrophysical bounds. The Imaging X-ray Polarimetry Explorer (IXPE) is now in the middle of a 5-year mission to measure cosmic X-ray polarization \cite{27}. So far, IXPE has identified polarized X-rays from magnetars with far higher X-ray fluxes \cite{28,29}, and the results have been utilized to constrain the axion scenario \cite{30}. It would be very intriguing if IXPE and future X-ray polarimeters could collect time-dependent polarization data, shedding light on the cause of their mystery hard X-ray excesses while also probing axion physics and axion-photon reconversion \cite{30a,30b,30bb} or axion-photon oscillation \cite{30c}.

We emphasize once more that the X-ray pulse structures caused by axions are generic for a large class of sufficiently isolated strange quark stars. It would be intriguing to use the Polyakov Loop Extension (PNJL Model) \cite{31} and Magnetic NJL (mNJL) \cite{32} framework to extract information about axion models from the timing data for much brighter sources, like magnetars, with both flux and polarization data available. This hasn't been done yet and could aid in resolving astrophysical backgrounds. On the other hand, if an axion with parameters capable of producing detectable X-ray signals is found, it might be used to investigate the characteristics of strange quark stars.

Our results emphasize the value of taking into account more intricate models when examining axion-photon conversions surrounding quark stars. Our models and observational methods can be improved as our knowledge of strange quark stars and axions grows. The information we collect will be crucial for refining our search strategies and eventually helping to solve the universe's dark matter riddles. Future searches for these elusive particles will be guided by our ongoing investigation of axion characteristics, interactions, and surroundings.

\section*{Acknowledgements}

This work was presented at the International Workshop ``Infinite and Finite Nuclear Matter'' (INFINUM-2025), 12-16 May 2025, BLTP, JINR, Dubna, Russian Federation. The author expresses gratitude to the director of BLTP, JINR Prof. Dmitri I. Kazakov, for his academic support.

\end{document}